\pdfoutput=1
\documentclass[sigconf,screen,nonacm]{acmart}

\usepackage{algpseudocode}
\usepackage{listings}
\usepackage{xcolor}
\usepackage{xspace}
\usepackage[capitalise]{cleveref}
\usepackage{enumitem}
\usepackage[draft]{minted}
\usepackage{microtype}

\newcommand{\tool}{\textsc{CompSuite}\xspace}
\newcommand{\runner}{\textsc{CompRunner}\xspace}

\newcommand{\code}[1]{\texttt{#1}}

\setminted[bash]{fontsize=\small,
  breaklines,
  linenos,
  xleftmargin=8pt,
  numbersep=5pt
}

\setminted[json]{fontsize=\small,
  breaklines,
  linenos,
  xleftmargin=8pt,
  numbersep=5pt
}

\makeatletter
\def\PYG@reset{\let\PYG@it=\relax \let\PYG@bf=\relax%
    \let\PYG@ul=\relax \let\PYG@tc=\relax%
    \let\PYG@bc=\relax \let\PYG@ff=\relax}
\def\PYG@tok#1{\csname PYG@tok@#1\endcsname}
\def\PYG@toks#1+{\ifx\relax#1\empty\else%
    \PYG@tok{#1}\expandafter\PYG@toks\fi}
\def\PYG@do#1{\PYG@bc{\PYG@tc{\PYG@ul{%
    \PYG@it{\PYG@bf{\PYG@ff{#1}}}}}}}
\def\PYG#1#2{\PYG@reset\PYG@toks#1+\relax+\PYG@do{#2}}

\@namedef{PYG@tok@w}{\def\PYG@tc##1{\textcolor[rgb]{0.73,0.73,0.73}{##1}}}
\@namedef{PYG@tok@c}{\let\PYG@it=\textit\def\PYG@tc##1{\textcolor[rgb]{0.24,0.48,0.48}{##1}}}
\@namedef{PYG@tok@cp}{\def\PYG@tc##1{\textcolor[rgb]{0.61,0.40,0.00}{##1}}}
\@namedef{PYG@tok@k}{\let\PYG@bf=\textbf\def\PYG@tc##1{\textcolor[rgb]{0.00,0.50,0.00}{##1}}}
\@namedef{PYG@tok@kp}{\def\PYG@tc##1{\textcolor[rgb]{0.00,0.50,0.00}{##1}}}
\@namedef{PYG@tok@kt}{\def\PYG@tc##1{\textcolor[rgb]{0.69,0.00,0.25}{##1}}}
\@namedef{PYG@tok@o}{\def\PYG@tc##1{\textcolor[rgb]{0.40,0.40,0.40}{##1}}}
\@namedef{PYG@tok@ow}{\let\PYG@bf=\textbf\def\PYG@tc##1{\textcolor[rgb]{0.67,0.13,1.00}{##1}}}
\@namedef{PYG@tok@nb}{\def\PYG@tc##1{\textcolor[rgb]{0.00,0.50,0.00}{##1}}}
\@namedef{PYG@tok@nf}{\def\PYG@tc##1{\textcolor[rgb]{0.00,0.00,1.00}{##1}}}
\@namedef{PYG@tok@nc}{\let\PYG@bf=\textbf\def\PYG@tc##1{\textcolor[rgb]{0.00,0.00,1.00}{##1}}}
\@namedef{PYG@tok@nn}{\let\PYG@bf=\textbf\def\PYG@tc##1{\textcolor[rgb]{0.00,0.00,1.00}{##1}}}
\@namedef{PYG@tok@ne}{\let\PYG@bf=\textbf\def\PYG@tc##1{\textcolor[rgb]{0.80,0.25,0.22}{##1}}}
\@namedef{PYG@tok@nv}{\def\PYG@tc##1{\textcolor[rgb]{0.10,0.09,0.49}{##1}}}
\@namedef{PYG@tok@no}{\def\PYG@tc##1{\textcolor[rgb]{0.53,0.00,0.00}{##1}}}
\@namedef{PYG@tok@nl}{\def\PYG@tc##1{\textcolor[rgb]{0.46,0.46,0.00}{##1}}}
\@namedef{PYG@tok@ni}{\let\PYG@bf=\textbf\def\PYG@tc##1{\textcolor[rgb]{0.44,0.44,0.44}{##1}}}
\@namedef{PYG@tok@na}{\def\PYG@tc##1{\textcolor[rgb]{0.41,0.47,0.13}{##1}}}
\@namedef{PYG@tok@nt}{\let\PYG@bf=\textbf\def\PYG@tc##1{\textcolor[rgb]{0.00,0.50,0.00}{##1}}}
\@namedef{PYG@tok@nd}{\def\PYG@tc##1{\textcolor[rgb]{0.67,0.13,1.00}{##1}}}
\@namedef{PYG@tok@s}{\def\PYG@tc##1{\textcolor[rgb]{0.73,0.13,0.13}{##1}}}
\@namedef{PYG@tok@sd}{\let\PYG@it=\textit\def\PYG@tc##1{\textcolor[rgb]{0.73,0.13,0.13}{##1}}}
\@namedef{PYG@tok@si}{\let\PYG@bf=\textbf\def\PYG@tc##1{\textcolor[rgb]{0.64,0.35,0.47}{##1}}}
\@namedef{PYG@tok@se}{\let\PYG@bf=\textbf\def\PYG@tc##1{\textcolor[rgb]{0.67,0.36,0.12}{##1}}}
\@namedef{PYG@tok@sr}{\def\PYG@tc##1{\textcolor[rgb]{0.64,0.35,0.47}{##1}}}
\@namedef{PYG@tok@ss}{\def\PYG@tc##1{\textcolor[rgb]{0.10,0.09,0.49}{##1}}}
\@namedef{PYG@tok@sx}{\def\PYG@tc##1{\textcolor[rgb]{0.00,0.50,0.00}{##1}}}
\@namedef{PYG@tok@m}{\def\PYG@tc##1{\textcolor[rgb]{0.40,0.40,0.40}{##1}}}
\@namedef{PYG@tok@gh}{\let\PYG@bf=\textbf\def\PYG@tc##1{\textcolor[rgb]{0.00,0.00,0.50}{##1}}}
\@namedef{PYG@tok@gu}{\let\PYG@bf=\textbf\def\PYG@tc##1{\textcolor[rgb]{0.50,0.00,0.50}{##1}}}
\@namedef{PYG@tok@gd}{\def\PYG@tc##1{\textcolor[rgb]{0.63,0.00,0.00}{##1}}}
\@namedef{PYG@tok@gi}{\def\PYG@tc##1{\textcolor[rgb]{0.00,0.52,0.00}{##1}}}
\@namedef{PYG@tok@gr}{\def\PYG@tc##1{\textcolor[rgb]{0.89,0.00,0.00}{##1}}}
\@namedef{PYG@tok@ge}{\let\PYG@it=\textit}
\@namedef{PYG@tok@gs}{\let\PYG@bf=\textbf}
\@namedef{PYG@tok@gp}{\let\PYG@bf=\textbf\def\PYG@tc##1{\textcolor[rgb]{0.00,0.00,0.50}{##1}}}
\@namedef{PYG@tok@go}{\def\PYG@tc##1{\textcolor[rgb]{0.44,0.44,0.44}{##1}}}
\@namedef{PYG@tok@gt}{\def\PYG@tc##1{\textcolor[rgb]{0.00,0.27,0.87}{##1}}}
\@namedef{PYG@tok@err}{\def\PYG@bc##1{{\setlength{\fboxsep}{\string -\fboxrule}\fcolorbox[rgb]{1.00,0.00,0.00}{1,1,1}{\strut ##1}}}}
\@namedef{PYG@tok@kc}{\let\PYG@bf=\textbf\def\PYG@tc##1{\textcolor[rgb]{0.00,0.50,0.00}{##1}}}
\@namedef{PYG@tok@kd}{\let\PYG@bf=\textbf\def\PYG@tc##1{\textcolor[rgb]{0.00,0.50,0.00}{##1}}}
\@namedef{PYG@tok@kn}{\let\PYG@bf=\textbf\def\PYG@tc##1{\textcolor[rgb]{0.00,0.50,0.00}{##1}}}
\@namedef{PYG@tok@kr}{\let\PYG@bf=\textbf\def\PYG@tc##1{\textcolor[rgb]{0.00,0.50,0.00}{##1}}}
\@namedef{PYG@tok@bp}{\def\PYG@tc##1{\textcolor[rgb]{0.00,0.50,0.00}{##1}}}
\@namedef{PYG@tok@fm}{\def\PYG@tc##1{\textcolor[rgb]{0.00,0.00,1.00}{##1}}}
\@namedef{PYG@tok@vc}{\def\PYG@tc##1{\textcolor[rgb]{0.10,0.09,0.49}{##1}}}
\@namedef{PYG@tok@vg}{\def\PYG@tc##1{\textcolor[rgb]{0.10,0.09,0.49}{##1}}}
\@namedef{PYG@tok@vi}{\def\PYG@tc##1{\textcolor[rgb]{0.10,0.09,0.49}{##1}}}
\@namedef{PYG@tok@vm}{\def\PYG@tc##1{\textcolor[rgb]{0.10,0.09,0.49}{##1}}}
\@namedef{PYG@tok@sa}{\def\PYG@tc##1{\textcolor[rgb]{0.73,0.13,0.13}{##1}}}
\@namedef{PYG@tok@sb}{\def\PYG@tc##1{\textcolor[rgb]{0.73,0.13,0.13}{##1}}}
\@namedef{PYG@tok@sc}{\def\PYG@tc##1{\textcolor[rgb]{0.73,0.13,0.13}{##1}}}
\@namedef{PYG@tok@dl}{\def\PYG@tc##1{\textcolor[rgb]{0.73,0.13,0.13}{##1}}}
\@namedef{PYG@tok@s2}{\def\PYG@tc##1{\textcolor[rgb]{0.73,0.13,0.13}{##1}}}
\@namedef{PYG@tok@sh}{\def\PYG@tc##1{\textcolor[rgb]{0.73,0.13,0.13}{##1}}}
\@namedef{PYG@tok@s1}{\def\PYG@tc##1{\textcolor[rgb]{0.73,0.13,0.13}{##1}}}
\@namedef{PYG@tok@mb}{\def\PYG@tc##1{\textcolor[rgb]{0.40,0.40,0.40}{##1}}}
\@namedef{PYG@tok@mf}{\def\PYG@tc##1{\textcolor[rgb]{0.40,0.40,0.40}{##1}}}
\@namedef{PYG@tok@mh}{\def\PYG@tc##1{\textcolor[rgb]{0.40,0.40,0.40}{##1}}}
\@namedef{PYG@tok@mi}{\def\PYG@tc##1{\textcolor[rgb]{0.40,0.40,0.40}{##1}}}
\@namedef{PYG@tok@il}{\def\PYG@tc##1{\textcolor[rgb]{0.40,0.40,0.40}{##1}}}
\@namedef{PYG@tok@mo}{\def\PYG@tc##1{\textcolor[rgb]{0.40,0.40,0.40}{##1}}}
\@namedef{PYG@tok@ch}{\let\PYG@it=\textit\def\PYG@tc##1{\textcolor[rgb]{0.24,0.48,0.48}{##1}}}
\@namedef{PYG@tok@cm}{\let\PYG@it=\textit\def\PYG@tc##1{\textcolor[rgb]{0.24,0.48,0.48}{##1}}}
\@namedef{PYG@tok@cpf}{\let\PYG@it=\textit\def\PYG@tc##1{\textcolor[rgb]{0.24,0.48,0.48}{##1}}}
\@namedef{PYG@tok@c1}{\let\PYG@it=\textit\def\PYG@tc##1{\textcolor[rgb]{0.24,0.48,0.48}{##1}}}
\@namedef{PYG@tok@cs}{\let\PYG@it=\textit\def\PYG@tc##1{\textcolor[rgb]{0.24,0.48,0.48}{##1}}}

\def\PYGZus{\char`\_}
\def\PYGZob{\char`\{}
\def\PYGZcb{\char`\}}

\def\PYGZsh{\char`\#}

\def\PYGZhy{\char`\-}

\def\PYGZdq{\char`\"}

\makeatother

\renewcommand{\paragraph}[1]{\vskip 0.05in \noindent {\bf #1.}}

\AtBeginDocument{%
  }

\setcopyright{acmcopyright}
\copyrightyear{2023}
\acmYear{2023}
\acmDOI{XXXXXXX.XXXXXXX}
\acmPrice{15.00}
\acmISBN{978-1-4503-XXXX-X/18/06}

\begin{document}

\title{\tool: A Dataset of Java Library Upgrade Incompatibility Issues}

\author{Xiufeng Xu}
\email{xiufeng001@e.ntu.edu.sg}
\affiliation{%
  \institution{Nanyang Technological University}
  \country{Singapore}
}

\author{Chenguang Zhu}
\email{cgzhu@utexas.edu}
\affiliation{%
  \institution{The University of Texas at Austin}
  \country{USA}
}

\author{Yi Li}
\email{yi_li@ntu.edu.sg}
\affiliation{%
  \institution{Nanyang Technological University}
  \country{Singapore}
}

\begin{abstract}
Modern software systems heavily rely on external libraries developed
by third-parties to ensure efficient development. However, frequent library upgrades can lead to
compatibility issues between the libraries and their client
systems. In this paper, we introduce \tool, a dataset that includes
123 real-world Java client-library pairs where upgrading the
library causes an incompatibility issue in the corresponding
client. Each incompatibility issue in \tool is associated with a test
case authored by the developers, which can be used to reproduce the
issue. The dataset also provides a command-line interface that simplifies the
execution and validation of each issue. With this infrastructure,
users can perform an inspection of any incompatibility issue with the push of a button, or
reproduce an issue step-by-step for a more detailed investigation.
We make \tool publicly available to promote open science.
We believe that various software analysis techniques, such as compatibility checking,
debugging, and regression test selection, can benefit from \tool.
\end{abstract}

\settopmatter{printfolios=true}

\maketitle

\section{Introduction}\label{sec:intro}


Modern software systems are becoming increasingly complex due to the
need for integrating various components developed by different teams
or organizations. These components are often subject to continuous
evolution, and as a result, ensuring that new
upgrades to third-party libraries do not cause any compatibility
issues with the existing software system is a challenging task. The
complexity of these systems and the number of dependencies involved
make it difficult to anticipate and identify incompatibilities that
may arise from updates to external components.
Incompatibility issues resulting from upgrades to external components can
compromise the reliability of software systems, potentially leading to
significant financial losses for the organizations that rely on these systems.





Many techniques have been proposed to address third-party library
compatibility issues, including regression
testing~\cite{mezzetti2018type,moller2019model}, static
analysis~\cite{foo2018efficient}, dependency conflict
detection~\cite{wang2021will}, and client-specific compatibility
checking~\cite{mora2018client,zhu2019framework}. These techniques address
library compatibility issues in different dimensions and have been
evaluated with their own isolated datasets.

An excellent dataset has the potential to serve as a valuable reference for future research in
this field. However, composing the dataset requires intricate
manual validation, e.g., confirming whether the cause of a test
failure is due to runtime exception, assertion violation, or other reasons.
Therefore, we propose \tool, the first incompatibility issue dataset focusing on
library behavioral incompatibility with concrete reproducible test cases.
Each test case is isolated and validated, enabling the direct manifestation of the
incompatibilities.

\tool comprises 123 real-world Java client-library pairs such that upgrading any library results in
incompatibility issues for the corresponding client.
Every incompatibility issue in \tool contains a test case created by developers, allowing for the
reproduction of the issue.
On top of this dataset, we also developed an automated command-line interface, which streamlines
all processes of the reproduction, such as downloading and compiling a projects, running target
tests and re-runing the tests after a library upgrade.
With this infrastructure, users may reproduce an incompatibility issue programmatically with
minimal efforts.


\paragraph{Contribution}
To summarize, we make the following contributions in this paper:%
\begin{enumerate}[leftmargin=*,topsep=2pt]
\item We construct a dataset, \tool, including 123
  reproducible, real-world client-library pairs that manifest
  incompatibility issues when upgrading the library. These data points
  originate from 88 clients and 104 libraries.
\item We created an automated command-line interface for the dataset.
With this interface, users are able to programmatically replicate an incompatibility issue
from the dataset with a single command.
The interface also offers separate commands for each step involved in the reproduction of
incompatibility issues.
\end{enumerate}


We envision that \tool to be used to evaluate various program analysis techniques, including
compatibility checking, module-level regression testing selection, and debugging techniques.
More detailed information can be found in \cref{sec:app}.

The dataset and tool are available at: \url{https://github.com/compsuite-team/compsuite}.



\section{Dataset Creation}\label{creation}

In this section, we outline the methodology and process employed to
create the \tool dataset.

\subsection{Subjects Selection}

\begin{table*}[ht]
    \setlength{\abovecaptionskip}{5pt}
    \caption{Details of clients and libraries included in \tool.}
  \small
    \begin{tabular}{lrrclr}
        \cline{1-3} \cline{5-6}
        \textbf{Client}         & \textbf{\#LoC} & \textbf{\#Star} &  & \textbf{Library}                            & \textbf{\#Maven Usage} \\ \cline{1-3} \cline{5-6}
        retrofit                & 29.7K          & 41.5K           &  & org.slf4j:slf4j-api                         & 62.5K                   \\
        apollo                  & 61.3K          & 28K             &  & com.google.guava:guava                      & 34.4K                   \\
        druid                   & 441.9K         & 26.8K           &  & org.scala-lang:scala-library                & 34K                     \\
        webmagic                & 17.4K          & 10.8K           &  & com.fasterxml.jackson.core:jackson-databind & 25.8K                   \\
        languagetool            & 171.2K         & 8.5K            &  & ch.qos.logback:logback-classic              & 25.5K                   \\ \cline{1-3} \cline{5-6}
        Other 83 clients (mean) & 371.6K         & 1.3K            &  & Other 99 libraries (mean)                   & 3.2K                    \\ \cline{1-3} \cline{5-6}
        All clients (mean)      & 358.7K         & 2.5K            &  & All libraries (mean)                        & 4.8K                    \\ \cline{1-3} \cline{5-6}
    \end{tabular}
    \label{client-library}
\end{table*}

To ensure the representativeness and reproducibility of the \tool dataset, we
focus on including high-quality and popular client projects and
libraries. The selection of client projects was sourced from
GitHub~\cite{github}, a widely recognized online community for hosting
open-source codebases. To ensure the inclusion of the most
popular projects, we systematically sorted all the available projects
in descending order based on their number of stars on GitHub and
selected the target clients from the top of the list. The selection of
libraries was sourced from Maven Central~\cite{mvnrepo}, which hosts
33.5M of Java libraries and their associated binaries, making it
a widely used repository of libraries for Java API and library
research~\cite{mostafa2017study,wu2016exploratory,qiu2016understanding,kula2014visualizing}. We include a library in the dataset only
if it has more than 100 usages (i.e., clients) on Maven
Central. Our selection criteria aimed to ensure the inclusion of
popular and widely used client projects and libraries in the dataset,
thereby maximizing its relevance and usefulness to the research
community.

Among the highly-rated client projects, our selection criteria focused
on those that use Maven~\cite{mvn} as their build systems, given its
widespread adoption and maturity. Maven provides a standardized
approach to managing Java projects and their dependencies, where each
library dependency in a Maven client project is represented as an item
in a \code{pom.xml} file, making it easy to identify and edit library
versions programmatically. Furthermore, Maven offers built-in
functionality for running unit tests and generating test reports,
which simplifies the identification and diagnosis of incompatibility
issues arising from test executions. Since Maven projects typically rely on Maven Central as their
centralized repository for hosting and downloading libraries, the process of obtaining and managing
libraries in our dataset is simplified.

\Cref{client-library} presents the top 5 client projects and libraries in the \tool dataset, ranked by popularity. For
each client project, we provide information on its lines of code (LoC)
and the number of stars it has received on GitHub, while for each
library, we include its number of usages by other projects from Maven Central.

In total, \tool comprises 123 incompatible client-library pairs. These pairs encompass 88 distinct
clients and 104 libraries altogether. On average, the affected clients have 2.5K stars on GitHub
and 358.7K lines of code, while incompatible libraries have 4.8K usages on Maven Central.
Thus, we believe that the incompatibility issues present in
the \tool dataset have a significant impact on a large number of
codebases and can affect many users of the libraries, either directly
or indirectly.

To ensure that all client projects in the dataset are executable and
the runs are reproducible, we performed a series of checks on each
project. First, we checked out the project to the version (SHA) at the time of
the dataset creation, which we refer to as the \emph{base version}.
Next, we ran the standard Maven project compilation command to verify if the
project compiles successfully. If the project fails to compile, we
excluded it from the dataset. Subsequently, we ran the standard Maven
test command to execute all the tests in the project, ensuring that
all tests pass on the base version. We excluded any project that fails
to pass tests at this stage. Finally, we only included the client
projects that successfully compile and pass all tests on the base version,
thereby ensuring that the dataset is only consist of projects which can
be executed and whose executions can be reproduced.

\subsection{Data Collection}\label{sec:data:collect}

\begin{figure*}[ht]
  \setlength{\abovecaptionskip}{5pt}
  \setlength{\belowcaptionskip}{0pt}
  \centering
  \includegraphics[width=0.9\textwidth]{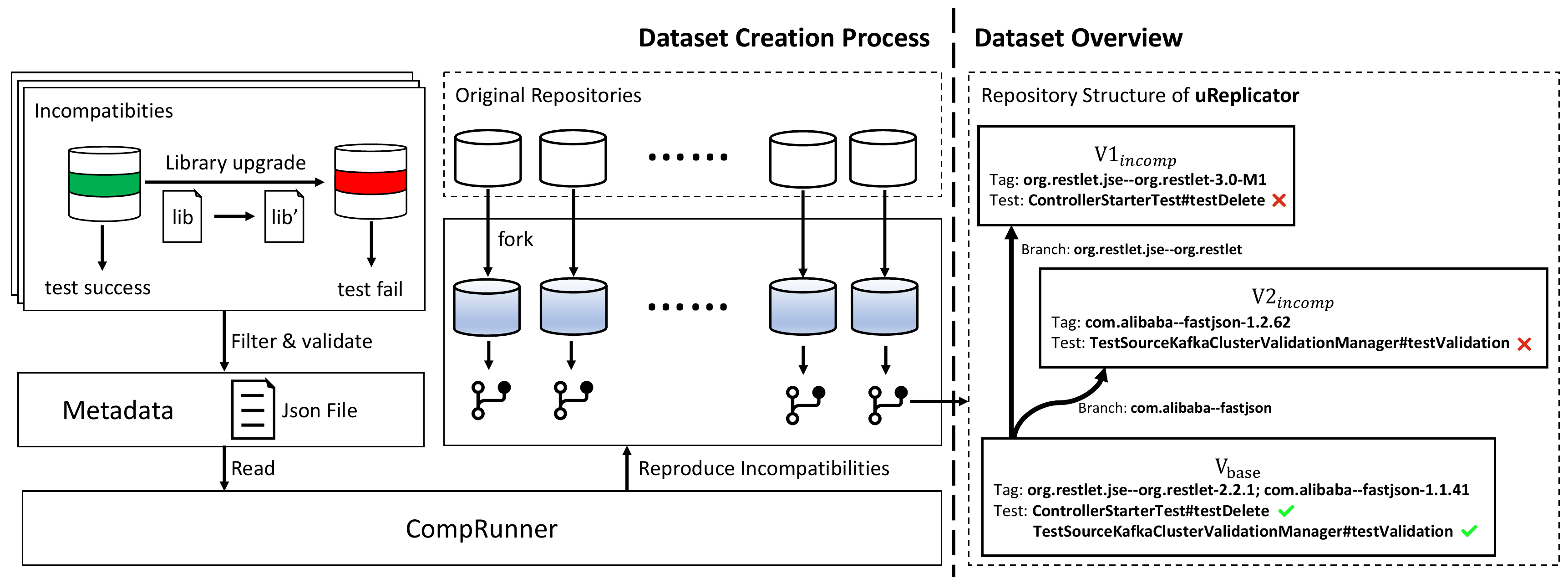}
  \caption{The architecture of \tool.}
  \label{fig:architecture}
\end{figure*}

We collected the data following the below procedures.
\Cref{fig:architecture} visualizes the overall architecture of \tool.
In the upper left portion of \cref{fig:architecture}, we
illustrate the approach taken by \tool to identify incompatibilities
between a client project and its dependent libraries. Specifically,
for each client project on its base version, we upgraded each of its
dependent libraries and tested if the upgrade caused any test
failures. Our intuition behind this approach is that since all the
tests in the client passed on the base version, if upgrading any
library causes a test failure, that library upgrade must have
introduced incompatibility issues. We refer to the test that flips
from passing to failing as an \emph{incompatibility-revealing test}.

To automatically upgrade the libraries and run the tests, we utilized
the Maven Versions Plugin~\cite{mvnplugin}. For a given client project, we
scanned its dependency list using this plugin to identify all the
libraries that had newer versions available on Maven Central. If a
library had a newer version, we marked it as upgradable. Next, for
each upgradable library, we used the plugin to upgrade it by updating
the \code{pom.xml} file to the most recent version on Maven
Central. We then re-executed the test suite of the client. If any
tests failed during this run, we marked the client-library pair as
having an incompatibility issue and marked the test as an
incompatibility-revealing test of this issue. It is crucial to note
that we only upgraded one library at a time to isolate failures caused
by different libraries. To ensure the accuracy and dependability of
the dataset, we carried out a manual verification process for each
identified incompatibility issue. In particular, we carefully examined
the test failure messages and reports to confirm that they were indeed
caused by the upgraded library. For each incompatible client-library
pair, we selected a single incompatibility-revealing test to be
included in the final dataset. In cases where a client-library pair
had multiple incompatibility issues, we chose the one that we deemed
most representative and easy to comprehend.

\begin{figure}
    \setlength{\abovecaptionskip}{5pt}
    \setlength{\belowcaptionskip}{0pt}
    \centering
\begin{Verbatim}[fontsize=\small,breaklines,linenos,xleftmargin=8pt,numbersep=5pt,commandchars=\\\{\}]
\PYG{p}{\PYGZob{}}
\PYG{+w}{  }\PYG{n+nt}{\PYGZdq{}id\PYGZdq{}}\PYG{p}{:}\PYG{+w}{ }\PYG{l+s+s2}{\PYGZdq{}i\PYGZhy{}49\PYGZdq{}}\PYG{p}{,}
\PYG{+w}{  }\PYG{n+nt}{\PYGZdq{}client\PYGZdq{}}\PYG{p}{:}\PYG{+w}{ }\PYG{l+s+s2}{\PYGZdq{}wasabi\PYGZdq{}}\PYG{p}{,}
\PYG{+w}{  }\PYG{n+nt}{\PYGZdq{}sha\PYGZdq{}}\PYG{p}{:}\PYG{+w}{ }\PYG{l+s+s2}{\PYGZdq{}9f2aa5f92e49c3844d787320e2d22e15317aa8e2\PYGZdq{}}\PYG{p}{,}
\PYG{+w}{  }\PYG{n+nt}{\PYGZdq{}url\PYGZdq{}}\PYG{p}{:}\PYG{+w}{ }\PYG{l+s+s2}{\PYGZdq{}https://github.com/intuit/wasabi\PYGZdq{}}\PYG{p}{,}
\PYG{+w}{  }\PYG{n+nt}{\PYGZdq{}lib\PYGZdq{}}\PYG{p}{:}\PYG{+w}{ }\PYG{l+s+s2}{\PYGZdq{}org.apache.httpcomponents:httpclient\PYGZdq{}}\PYG{p}{,}
\PYG{+w}{  }\PYG{n+nt}{\PYGZdq{}old\PYGZdq{}}\PYG{p}{:}\PYG{+w}{ }\PYG{l+s+s2}{\PYGZdq{}4.5.1\PYGZdq{}}\PYG{p}{,}
\PYG{+w}{  }\PYG{n+nt}{\PYGZdq{}new\PYGZdq{}}\PYG{p}{:}\PYG{+w}{ }\PYG{l+s+s2}{\PYGZdq{}4.5.10\PYGZdq{}}\PYG{p}{,}
\PYG{+w}{  }\PYG{n+nt}{\PYGZdq{}test\PYGZdq{}}\PYG{p}{:}\PYG{+w}{ }\PYG{l+s+s2}{\PYGZdq{}DefaultRestEndPointTest\PYGZsh{}testGetRestEndPointURI\PYGZdq{}}\PYG{p}{,}
\PYG{+w}{  }\PYG{n+nt}{\PYGZdq{}submodule\PYGZdq{}}\PYG{p}{:}\PYG{+w}{ }\PYG{l+s+s2}{\PYGZdq{}modules/export\PYGZdq{}}\PYG{p}{,}
\PYG{+w}{  }\PYG{n+nt}{\PYGZdq{}test\PYGZus{}cmd\PYGZdq{}}\PYG{p}{:}\PYG{+w}{ }\PYG{l+s+s2}{\PYGZdq{}mvn org.apache.maven.plugins:maven\PYGZhy{}surefire\PYGZhy{}plugin:2.20:test \PYGZhy{}fn \PYGZhy{}Drat.ignoreErrors=true \PYGZhy{}DtrimStackTrace=false \PYGZhy{}Dtest=DefaultRestEndPointTest\PYGZsh{}testGetRestEndPointURI\PYGZdq{}}
\PYG{p}{\PYGZcb{}}
\end{Verbatim}
\caption{The data schema of \tool}\label{fig:json}
\end{figure}

Finally, we persisted the metadata of all the selected incompatibility
issues in a collection of \code{json} files.
\Cref{fig:json} presents the metadata of an incompatibility
issue in the \tool dataset. The data schema includes the ID of the
issue, client project name, SHA of the client base version, URL of the
client project, library name, versions of the old and new libraries,
the name of the incompatibility-revealing test, the submodule
containing the incompatibility-revealing test, and the command to run
the test. The majority of the information is
self-explanatory. However, it is worth noting that the old version of
the library is the one utilized at the base version of the client,
while the new version is the most recent version found on Maven
Central that triggers the incompatibility when upgrading, as described in
\cref{sec:data:collect}.

\section{Dataset Usage}\label{sec:usage}

In this section, we provide instructions on the usage of our dataset.

\subsection{Exploring an Incompatibility Issue}\label{sec:isolate}
To ensure the reproducibility of incompatibility issues and to
facilitate the demonstration of such issues, we have annotated
checkpoints in the version histories of the client projects and
provided tags that guide users to explore any incompatibility issues
present in the \tool dataset.

As illustrated on the right-hand side of \cref{fig:architecture}, our
approach to handling incompatible client-library pairs involved
creating a fork~\cite{gitfork} of the original client project for each
identified pair, while preserving all code and version history
information. To mark the base version of the project, we utilized the
\code{git tag}~\cite{gittag} command, designating it as $V_{base}$. Subsequently, we developed a
patch to upgrade the library from its old version to its new version,
a simple process that can be accomplished with a single line change in
the \code{pom.xml} file for Maven projects. This patch was then
applied to the $V_{base}$ version, resulting in a new version that we
identified as $V_{incomp}$. Notably, the only difference between
$V_{base}$ and $V_{incomp}$ lies in the library version used: the old
(compatible) version is utilized on $V_{base}$ while the new
(incompatible) version is utilized on $V_{incomp}$. For instance, in
\cref{fig:architecture}, the client project employs version 2.2.1 of
the \code{org.restlet.jse-org.restlet} library on its $V_{base}$ and
version 3.0-M1 on its $V_{incomp}$. In cases where multiple libraries
exhibit incompatibility issues in the client project, we not only create different branches for each library with its name, but also generate a
$V_{incomp}$ version tag for each, with accompanying annotations that
denote the corresponding library name and version, as depicted in
\cref{fig:architecture}.

The $V_{incomp}$ tag for each client-library pair also specifies the
specific test that can reveal the incompatibility issue during its
run. Following Maven's convention, the test name is formatted as
\code{TestClassName\#testMethodName}. By simply copying the text from
the tag, users can easily run the incompatibility-revealing test on
the $V_{incomp}$ version and observe the incompatibility issue. On the
$V_{base}$ version, all tests should pass. This design aims to
simplify the usage of \tool and make it more accessible and
user-friendly.

Using the forked client repositories and version tags provided in the
\tool dataset, users can easily reproduce any incompatibility issue by
checking out to $V_{incomp}$ and running the corresponding
incompatibility-revealing test. To compare the behaviors of the client
with compatible and incompatible library versions, users can run the
incompatibility-revealing test on both $V_{base}$ and $V_{incomp}$ and
compare the test outcomes. This allows for a clear understanding of
the impact of the library upgrade on the client behaviors.

\subsection{\runner: An Automated Tool for Reproducing Incompatibility Issues}



We further developed an automated tool, named \runner, which is a part of
\tool. With \runner, users can easily reproduce and investigate any
incompatibility issue in a one-click manner by providing the issue ID
as input.

We offer an option which enables users to reproduce an
incompatibility issue end-to-end with a single command as is shown below.
The command outputs and saves all intermediate results and logs for future
reference.


\begin{Verbatim}[fontsize=\small,breaklines,linenos,xleftmargin=8pt,numbersep=5pt,commandchars=\\\{\}]
python\PYG{+w}{ }main.py\PYG{+w}{ }\PYGZhy{}\PYGZhy{}incompat\PYG{+w}{ }i\PYGZhy{}56
\end{Verbatim}

When \runner runs, it clones the client project from our forked
code repository and saves it in the output directory (which is configurable).
Then, it checks out to the base version, compiles the code, and runs the
incompatibility-revealing test. Next, it upgrades the library to the
new version, reruns the incompatibility-revealing test, and reports
any failure information to the user.

We also provide a set of commands that break down the entire cycle of incompatibility
exploration into separate steps:


\begin{Verbatim}[fontsize=\small,breaklines,linenos,xleftmargin=8pt,numbersep=5pt,commandchars=\\\{\}]
python main.py \PYGZhy{}\PYGZhy{}download i\PYGZhy{}56
python main.py \PYGZhy{}\PYGZhy{}compile i\PYGZhy{}56
python main.py \PYGZhy{}\PYGZhy{}testold i\PYGZhy{}56
python main.py \PYGZhy{}\PYGZhy{}testnew i\PYGZhy{}56
\end{Verbatim}


We provide several other \runner commands for users to inspect
different aspects of the incompatibility issues from the \tool dataset. A
complete list of these commands can be found on \tool's website at
\url{https://github.com/compsuite-team/compsuite}.


\section{Application Scenarios}\label{sec:app}

We anticipate that both researchers and practitioners can benefit from
\tool to facilitate their investigations and research on errors
and test failures induced by library upgrades. \tool supports the evaluation of
various program analysis techniques, such as software upgrade compatibility checking, debugging,
and module-level regression test selection techniques.

As an overview, authors of compatibility checkers and detectors
may use \tool as a benchmark to evaluate the performance of their
techniques against other baseline approaches. Furthermore, authors of
debugging techniques can utilize \tool as a dataset of compatibility
bugs, where each bug corresponds to a test case that verifies the
existence or absence of the bug. Finally, authors of module-level
regression test selection techniques can use \tool to assess the
safety of their approaches. A safe module-level RTS technique should
select all the corresponding incompatibility-revealing test cases when
the library changes.

We detail the three usage scenarios as follows.
\begin{itemize}[leftmargin=*,topsep=2pt]
    \item \textbf{Compatibility Checkers and Detectors.}
      The existing techniques for compatibility checking and detection
      in Java can be categorized into three groups: i) Techniques for
      detecting API incompatibility that focus on detecting
      API-breaking changes, such as renaming of code entities and
      changes in parameter types~\cite{li2018cid,he2018understanding,scalabrino2019data,wei2019pivot,huang2018understanding}. ii)
      Techniques for detecting behavioral incompatibility that focus
      on identifying behavioral differences that cause test failures
      when a library is upgraded in a client, such as changes in
      program states~\cite{zhuclient,mora2018client}. iii) Techniques for
      detecting dependency conflicts~\cite{wang2021will,wang2018dependency,wang2019could},
      which aim to identify library APIs that exhibit inconsistent
      semantics between libraries due to class path shading. We
      believe that developers of techniques in the first two
      categories can use \tool as a benchmark to evaluate their tools'
      performance, such as precision and recall. They can run their
      tools on the \tool dataset and compare the results with the
      incompatibility issues present in the dataset. On the other
      hand, developers of techniques for detecting dependency
      conflicts can slightly modify \tool's dataset by placing both
      old and new libraries on the class path, running library
      conflict detection, and checking if the issues can be detected.

    \item \textbf{Module-Level Regression Test Selection.}  Regression
      test selection (RTS) is a technique that aims to reduce the cost
      of regression testing by selecting a subset of tests that may
      change the behavior due to code changes on each program
      version~\cite{gligoric2015practical,legunsen2016extensive,zhang2018hybrid,zhu2019framework}. Module-level
      RTS focuses on selecting the affected client tests when a
      dependent library is updated~\cite{gyori2018evaluating}. The developers of
      module-level RTS techniques can evaluate the safety of their
      tools using \tool. For each client-library pair in \tool, a
      module-level RTS tool should select all the corresponding
      incompatibility-revealing tests when upgrading the library from
      the old version to the new version.

    \item \textbf{Debugging.}  The existing debugging techniques for
      Java, include symbolic execution~\cite{baldoni2018survey}, delta
      debugging~\cite{zeller1999yesterday}, fault localization~\cite{wong2016survey}, etc. These techniques aim to identify the root
      cause of errors or failures in software. Developers of debugging
      techniques can use \tool as a dataset of compatibility bugs,
      where each compatibility bug corresponds to a test case that
      checks the presence or absence of the bug. They can use \tool to
      evaluate their techniques' ability to perform root cause
      analysis by trying to identify the corresponding library change
      that caused the compatibility issue.
\end{itemize}

\section{related work}\label{sec:related}

To cater to the requirements of various research endeavors, numerous outstanding datasets have been
made available to date. Just et al.~\cite{just2014defects4j} introduced Defects4J, a database
supplies actual bugs, fixed program versions, and corresponding test suites. Bui et
al.~\cite{bui2022vul4j} introduced Vul4J focusing on Java vulnerabilities. Jezek et al.
\cite{jezek2017api} released their dataset of compatibility issues arising from program evolution.
There are also many datasets cater for other research domains and
ecosystems~\cite{zhu2017dataset,wei2016taming,nielebock2021androidcompass}.

Distinct from the previously discussed datasets, \tool is the first dataset emphasizes the
incompatibility issues caused by Java library behavior changes. This type of issues are prevalent
and difficult to detect. Additionally, Our developed automated tools also have the capability to assist researchers in swiftly reproducing issues. We believe that a dataset targeting the library upgrade incompatibility
issue will contribute to the advancement of the associated technologies.

\section{Conclusion}\label{sec:conclude}

This paper presents \tool, a dataset containing 123 real-world Java client-library pairs where
library upgrades cause compatibility issues in the corresponding clients. On top of it, we also
developed a command-line interface, \runner, which allows users to quickly check incompatibility
issues with a single command or reproduce an incompability programmatically for in-depth analysis.
We believe that various program analysis techniques, such as library compatibility checking,
debugging, and regression test selection, may benefit from our dataset.

\bibliographystyle{acm}
\bibliography{main}

\end{document}